\documentstyle[10pt]{article}
\begin{document}

\def\beq{\begin{equation}}
\def\eeq{\end{equation}}
\def\bce{\begin{center}}
\def\ece{\end{center}}
\def\ben{\begin{enumerate}}
\def\een{\end{enumerate}}
\def\ul{\underline}
\def\ni{\noindent}
\def\nn{\nonumber}
\def\bs{\bigskip}
\def\ms{\medskip}
\def\wt{\widetilde}
\def\wh{\widehat}
\def\Tr{\mbox{Tr}\ }

\hfill DFTUZ/95/14

\hfill March, 1995

\vspace*{3mm}

\begin{center}

{\Large \bf Conformal Transformation in Gravity}

\vspace{30mm}

\medskip

{\sc Ilya L. Shapiro} \footnote{E-mail: shapiro@dftuz.unizar.es}\\
Departamento de Fisica Teorica, Universidad de Zaragoza,
50009, Zaragoza, Spain\\
and\\
Tomsk Pedagogical Institute, 634041 Tomsk, Russia.

\vskip 10mm

{\sc Hiroyuki Takata}\footnote{ E-mail:
 takata@theory.kek.jp}\\
Department of Physics, Hiroshima University, Higashi-Hiroshima 739,
Japan\\
and\\
National Laboratory for High Energy Physics (KEK),Tsukuba,Ibaraki 305,Japan.

\vskip 5mm
\vspace{25mm}

{\bf Abstract}

\end{center}
The  conformal transformation in the Einstein - Hilbert action leads to
a new frame where  an extra scalar  degree of freedom is compensated by
the local conformal-like symmetry.  We write down a most general action
resulting from such transformation and show that it covers both general
relativity  and  conformally  coupled  to  gravity scalar  field as the
particular cases. On quantum level the equivalence between the different
frames is disturbed by the loop corrections. New conformal-like symmetry
in anomalous and, as a result, the theory is not finite on shell at the
one-loop order.

\vspace{4mm}


\vspace{5mm}

\noindent{\large \bf 1 Introduction}
\vskip 1mm

The frame dependence of the gravity theories is an important object for
study on both classical and quantum levels. On classical level the theory
may manifest a different physical properties in a different frames that
leads to the nontrivial problems related with the "correct choice" of
the field variables (see, for example,  \cite{1,2,3,DEF,DN,bar,hj,4}
and the last work for a more complete list of references on the subject).
On the other hand the study of different frames enables one to explore
the relation between the different physical theories and thus generate the
new exact solutions \cite{5,6,7}.
In this paper we construct and study the general action of the theory
which is conformally equivalent to General Relativity with (or without)
cosmological constant. The theory
under consideration depends on the metric and also on the scalar
field, whereas an extra scalar degree of freedom is compensated by
additional conformal-like symmetry. In fact we exchange the theory with the
Einstein - Hilbert action to the much more complicated but equivalent
theory with the action depending on some arbitrary function of the
scalar field. The form of the symmetry transformation depends on the form
of this function.

On quantum level the equivalence between two formulations is disturbed by
anomaly which is essential and manifest itself at the level of the
one-loop divergences already.
In a framework of Einstein gravity the one loop counterterms vanish
on classical mass shell and the theory is finite \cite{hove}. Indeed
this property does not hold if the matter fields are incorporated
\cite{dene} or if the two-loop effects are taken into account \cite{GS}.
In our  new conformal frame the one-loop $S$-matrix is not finite
because of  anomaly.  This fact can be interpreted
as the noninvariance of the measure of path integral with respect to
(generalized) conformal transformations. Earlier the similar objection
have been made in a quantum conformal (Weyl) gravity \cite{frts,a}
which is power counting renormalizable. In the case of Weyl gravity the
lack of renormalizability is caused by anomaly which affects even the
one-loop divergences. The difference is that here we lack of the
one-loop on shell renormalizability which is lost in a new frame.

The paper is organized as follows. In section 2 we consider
the conformal transformation in general dilaton gravity, establish
the new frame for General Relativity and discuss the corresponding
conformal-like symmetry.  For the sake of convenience of the reader
we rewrite a small part of our previous paper \cite{spec} in the
beginning of this section. We remark that the conformal
transformation in a dilaton-gravity models have been studied in
\cite{deser,3}. Some special case of the nonconformal theory
with an interesting duality symmetry is discussed in an Appendix.
Throughout the paper we prefer to deal with the action rather than
with the equations of motion because it is more suitable for quantum
consideration which follows in section 3. In this section we derive
the one-loop counterterms for the theory in an arbitrary conformal
frame and find that these counterterms are not conformal invariant
and do not vanish on shell. The last section is conclusion.

\vspace{5mm}

\noindent{\large \bf 2 General theory with conformal-like symmetry}
\vskip 1mm

Let us start our study with the general  four dimensional metric-dilaton
model including the second derivative terms only. The most general
theory of this type is described by the action
\beq
S= \int d^4x \sqrt{-g}\; \{ A(\phi)g^{\mu\nu}\partial_{\mu}\phi
\partial_{\nu}\phi + B(\phi)R + C(\phi) \}
                                                    \label{0.1}
\eeq
that covers all special cases including the string inspired action,
coupled with gravity scalar field and others.
Now, following \cite{3,spec} we consider the
simple particular case
of the general action (\ref{0.1}).\footnote{One can consider the theory
(\ref{1.1}) as
the four-dimensional analog of the Jakiw-Teitelboim model in two dimensions.}
\beq
S= \int d^4x \sqrt{-g'} \; \{ R'\Phi+ V(\Phi) \}
                                                    \label{1.1}
\eeq
Here the curvature $R'$ corresponds to the metric $g'_{\mu\nu}$ and
$g'= \det (g'_{\mu\nu})$. Transform this action to the new variables
$g_{\mu\nu}$ and $\phi$ according to
\beq
g'_{\mu\nu}=g_{\mu\nu}e^{2\sigma(\phi)},\;\;\;\;\;\;\;\;\;\;\;
\Phi=\Phi(\phi)                                     \label{1.2}
\eeq
where $\sigma(\phi)$ and $\Phi(\phi)$ are arbitrary functions of
$\phi$. In a new variables the action becomes:
\beq
S= \int d^4x \sqrt{-g}\; \left\{ \Phi(\phi)Re^{2\sigma(\phi)}+
6(\nabla\phi)^2e^{2\sigma(\phi)}[\Phi\sigma'+\Phi']\sigma'
+ V(\Phi(\phi))e^{4\sigma(\phi)} \right\}           \label{1.3}
\eeq
Therefore we are able to transform the particular action (\ref{1.1})
to the general form (\ref{0.1}) with
\beq
A(\phi)=6e^{2\sigma(\phi)}[\Phi\sigma_1+\Phi_1]\sigma_1,\;\;\;\;\;\;\;\;\;\;\;
B(\phi)=\Phi(\phi) e^{2\sigma}                     \label{1.4}
\eeq
Here and below the lower numerical index shows
the order of derivative with respect to $\phi$. For instance,
\beq
B_1 = \frac{dB}{d\phi},\ \ \ \ \ A_2  = \frac{d^2 B}{d\phi^2},\ \ \ \
\sigma_1 = \frac{d\sigma}{d\phi}, \ \ \ etc.   \label{1.6}
\eeq
It is possible to
find the form of $\sigma(\phi)$ and $\Phi(\phi)$ that corresponds to the
given $A(\phi)$ and $B(\phi)$. In this case $\sigma(\phi)$
and $\Phi(\phi)$ obey the equations
\beq
 A=6B_1\sigma_1-6B(\sigma_1)^2,
\ \  \ \ \ \ \ \ \  \Phi = B e^{-2\sigma}                     \label{1.7}
\eeq
Substituting (\ref{1.7}) into (\ref{1.3}) we find that in a new variables the
action has the form
\beq
S= \int d^4x \sqrt{-g} \;\{ A(\phi)g^{\mu\nu}\partial_{\mu}\phi
\partial_{\nu}\phi + B(\phi)R + \left( \frac{B}{\Phi}\right)^2
V(\Phi(\phi))\}                                          \label{1.8}
\eeq
where the last term is nothing but $C(\phi)$ from (\ref{0.1}).

It is easy to see that the above transformations lead to some restrictions
on the functions $A(\phi)$ and $B(\phi)$. All the consideration
has to be modified if $\Phi=const$, that is the  case when (\ref{1.1})
is the Einstein-Hilbert action with cosmological constant.
One can rewrite this condition in terms of
$A(\phi)$ and $B(\phi)$. Note that
\beq
 2AB -3(B_1)^2 = -
3\;(\frac{d\Phi}{d\phi})^2 \;e^{4\sigma(\phi)}   \label{1.5}
\eeq
Hence it is clear that the theories with
 $\;2AB -3(B_1)^2 =0\;$
 qualitatively differs
from the other ones. Below we deal with the theories of this class.
The only exception is the Appendix.

One can easily see that the Hilbert-Einstein action and  the
conformally coupled with gravity scalar field satisfy the condition
(\ref{1.5}) with $B = const, A = 0, C = const$ and
$A=\frac12, B=\frac{1}{12}\phi^2, C=\lambda\phi^4$ correspondingly.
Therefore these two theories belongs to the special class of
(\ref{0.1}) which we are dealing with. Thus we have found that the
theory of conformal scalar field and General Relativity are related
with each other by the conformal transformation of the metric. The
conformal symmetry compensate an extra degree of freedom which exists
in the conformal scalar theory.

It is possible to show that this situation is typical.
In fact all the models
(\ref{0.1}) with  $\;2AB -3(B_1)^2 =0\;$
 are related with each other by the conformal
transformation of the metric. Next, all of them besides the degenerate
case of General Relativity have some extra conformal-like symmetry that
is called to compensate an extra degree of freedom related with scalar
field. It is useful to formulate this as two theorems.
\vskip 1mm
{\bf Theorem 1.}
If two sets of smooth functions $A(\phi), B(\phi), C(\phi)$ and
${\bar A}(\phi), {\bar B}(\phi), {\bar C}(\phi)$ satisfy the conditions
\beq
2AB -3(B_1)^2 =0,\;\;\;\;\;\;\;\;\;\;\;
C=\lambda B^2                                            \label{1.10}
\eeq
\beq
2{\bar A}{\bar B} - 3({\bar B}_1)^2 =0, \;\;\;\;\;\;\;\;\;\;\;
{\bar C} = \lambda {\bar B}^2                           \label{1.101}
\eeq
then the corresponding models (\ref{0.1}) are linked by the conformal
transformation of the metric
${\bar g}_{\mu\nu} = g_{\mu\nu} e^{2\sigma(\phi)}$
where the parameter of transformation  $\sigma$ depends on the ratio
$\frac{{\bar B}(\phi)}{B(\phi)}$.
\vskip 1mm
{\bf Theorem 2.}
The action of any theory from the previous Theorem is invariant under
the transformation which consists in
 an arbitrary reparametrization ${\bar \phi} =
{\bar \phi} (\phi)$ and the conformal transformation
${\bar g}_{\mu\nu} = g_{\mu\nu} e^{2\sigma(\phi)}$ with
\beq
\sigma(\phi)  = - \frac12\;
\ln \left[ \frac{B({\bar \phi}(\phi)}{B({\phi})} \right]  \label{c}
\eeq

{\bf Proof.}  To cover both theorems let us consider
the action (\ref{0.1}), (\ref{1.10}) with and make the conformal
transformation of the metric with arbitrary $\sigma(\phi)$
and, simultaneously, an arbitrary reparametrization
of the scalar field ${\bar \phi} =
{\bar \phi} (\phi)$. \footnote{Indeed we suppose that all the functions
are mathematically acceptable for our manipulations.}
The straightforward calculation gives the new action with the
functions ${\bar A}(\phi), {\bar B}(\phi), {\bar C}(\phi)$
which satisfy the conditions
$$
B(\phi) = {\bar B}({\bar \phi}(\phi))
e^{2\sigma(\phi)},\;\;\;\;\;\;
C(\phi) = {\bar C}({\bar \phi}(\phi)) e^{4\sigma(\phi)}
$$
\beq
A(\phi) = \left\{
{\bar A}({\bar \phi}(\phi))
\left[ {\bar \phi}(\phi) \right]^2
+ 6{\bar B}({\bar \phi}(\phi)) \left[ \sigma_1(\phi) \right]^2
+ 6 \frac{d{\bar B}({\bar \phi}(\phi))}{d{\bar \phi}(\phi))}
\left[ \sigma_1(\phi) \right] \left[ {\bar \phi}_1(\phi) \right]
\right\} e^{2\sigma(\phi)}                           \label{1.11}
\eeq
The first theorem results from (\ref{1.11}) with
$ {\bar \phi}(\phi) = \phi $.
It is easy to see that if the functions
${\bar A}(\phi), {\bar B}(\phi),
{\bar C}(\phi)$ satisfy the
conditions (\ref{1.101}) then the functions $A(\phi), B(\phi), C(\phi)$
satisfy (\ref{1.10}). Therefore for any given ${\bar B}(\phi)$ and
$B(\phi)$ one can take $\sigma(\phi)=
\frac12\;\ln \left( \frac{ B(\phi)}{{\bar B}(\phi)} \right)$ that complete
the proof.

Now we suppose that $ {\bar \phi}(\phi)$
is arbitrary function and
require the action to be invariant under the reparametrization plus
conformal transformation, that evidently gives (\ref{c}).

It is interesting to consider a very simple example of that how Theorem 2
works. If ${\bar \phi}(\phi) = \phi e^{-\rho(x)}$ where $\rho(x)$ is
some arbitrary function of the spacetime variables,
and ${\bar B}({\bar \phi}) = \frac{1}{12} {\bar \phi}^2$ then
we find $B(\phi) = \frac{1}{12}\phi^2 e^{-2\rho(x)}$ from what follows
$\sigma(\phi)=\rho(x)$. Therefore for the particular case of the
conformally coupled with gravity scalar field the symmetry established
in the Theorem 2 is nothing but usual conformal symmetry.

And so we have found that the theories (\ref{0.1}) are
distinguished (one can say labeled) by the
form of the functions $A, B, C$. One can imagine  some three
dimensional space where the functions $A, B, C$ play the roles of
coordinates. One-dimensional line in this space is composed by the
theories which satisfy (\ref{1.10}). All of them are related with each other
by the conformal transformation described in Theorem 1. Next, all of them
with one exception of General Relativity possess an extra conformal-like
symmetry according to the Theorem 2. Since all the models of this
class are conformally equivalent to  General Relativity they all have
the same physical content and can be regarded as different frames for
description of gravity. In particular, it is possible to obtain exact
solutions for any of such theories with the use of conformal transformation
in any of the known solutions of Einstein Gravity with (or without)
cosmological constant.

It is interesting to consider the possibility of the soft breaking of
the new conformal-like symmetry. To do this it is useful to construct it's
Noether identity. Taking into
account the transformation rules from the Theorem 2, one can easily
derive such identity in the form
\beq
B_1(\phi)\; g_{\mu\nu} \;\frac{\delta S}{\delta g_{\mu\nu}} -
B(\phi)\;\frac{\delta S}{\delta \phi} = 0               \label{1.13}
\eeq
where the factor $-\frac{2B}{\phi B_1}$ stands for the
 conformal weight of the scalar field $\phi$  which depends on the form
of the function $B(\phi)$.
It is an analog (one can say generalization) of the
ordinary conformal weight "$-1$"of the field $\phi$.
The eq. (\ref{1.13}) is the operator form
of the symmetry transformation established by Theorem 2.
It shows that in the presence of the
conformal-like symmetry
the equations of motion are linearly dependent.

The soft breaking of the symmetry means that the functions
$A$ and $B$ satisfy the symmetry condition (\ref{1.10})
whereas the restrictions on the potential term $C$
are not imposed. It turns out that only the invariant form
of $C(\phi)$ is consistent with the equations of motion.
 From (\ref{1.13}) follows that the kinetic (that is $A$ and $B$ dependent)
parts of the equations of motion are linearly dependent. Substituting an
arbitrary function $C(\phi)$ into (\ref{1.13}) we arrive at the
differential equation for $C:\;\;$
$C_1(\phi)B(\phi) = 2C(\phi)B_1(\phi)$ that leads to $C(\phi) = B^2(\phi)
\cdot const$. Thus in a pure theory without matter only the symmetric
form of $C(\phi)$ is consistent with the equations of motion and any soft
symmetry breaking is forbidden.
For the standard conformal symmetry this was pointed out by Ng \cite{ng}.
One can easily check that this statement
is correct even if we add the action of matter, if this matter does not
depend on the field $\phi$. Thus if we consider the theory (\ref{0.1})
with $2AB-(B_1)^2=0$ then only in the case $C(\phi)=\lambda B^2(\phi),\;\;
\lambda = const$ there can exist any solutions of the dynamical equations.

\vspace{5mm}

\noindent{\large \bf 3 Divergences and conformal anomaly}
\vskip 1mm

The next purpose of the present paper is to investigate the symmetric version
of the theory (\ref{0.1}), (\ref{1.10})
on quantum level. According to the Theorems 1,2
all the models which possess an extra conformal-like symmetry
are conformally equivalent to General Relativity with cosmological constant.
The different versions of the symmetric conformally equivalent models
can be labeled by the values of function $B(\phi)$ and constant $\lambda$,
as $A(\phi) = \frac{3B_1^2(\phi)}{2B(\phi)}$ and
$C(\phi) = \lambda B^2(\phi)$.
It is useful to denote the action of the symmetric
theory as $S_{B(\phi),\lambda}$.
General Relativity with cosmological constant corresponds to
$S_{\gamma,\lambda}$
where "$- \gamma$" is an inverse Newtonian constant.
Thus our calculation of the one-loop divergences in general
$S_{B(\phi),\lambda}$
theory may be considered as the calculation for the special case of
$S_{\gamma,\lambda}$
in a conformally transformed quantum variables.
For the sake of brevity we shall denote as $S_{B(\phi),\lambda}$
only the action of the theory with a nonconstant $B(\phi)$
and preserve the notation $S_{\gamma,\lambda}$ for General Relativity.

Before starting the calculations let us say some words about what result
we can expect.  The general theory $S_{B(\phi),\lambda}$ differs from
$S_{\gamma,\lambda}$ in one respect. The first one has one more field
variable that is compensated by an extra conformal-like symmetry.
On classical level both theories are equivalent. However on quantum
level the equivalence may be broken by anomaly which can violate
the symmetry. The example of Weyl conformal gravity has learned us
that in quantum gravity the conformal anomaly can affect the one-loop
divergences already \cite{frts,a}. The Weyl (conformal)
gravity is higher derivative
theory where the renormalizability is disturbed only by the conformal
anomaly. Our purpose here is to check whether the conformal anomaly
exists for the second derivative theory under consideration. Since
the source of anomaly is the noninvariance of the measure of the
path integral \footnote{In fact this noninvariance is caused by
the UV divergences because the regularization scheme doesn't
preserve both diffeomorphism and conformal invariance. It is quite
possible that the IR effects can also be relevant, but this is not
clear yet.} our study concerns the divergent anomalous
part of the Jacobian of an arbitrary conformal transformation from
$S_{\gamma,\lambda}$  to $S_{B(\phi),\lambda}$.

The simple consideration based on power counting
shows that the theory $S_{B(\phi),\lambda}$
is nonrenormalizable just as General Relativity.
The one-loop counterterms contain the terms of fourth order in
derivatives. The most general action of this type has the form
\cite{ejos,spec}:
$$
\Gamma_{div}^{1-loop}=\frac{1}{16 \pi^2 (n-4)}
\int d^4x\sqrt{-g}
[ c_w C^2 + c_r R^2 + c_4 R(\nabla \phi)^2 + c_5 R(\Box \phi )+
c_6 R^{\mu \nu}(\nabla_{\mu} \phi)(\nabla_\nu \phi) +
$$
\beq
+c_7 R +
c_8 (\nabla \phi)^4 + c_9 (\nabla \phi)^2(\Box \phi)
+ c_{10} (\Box \phi)^2 + c_{11}  (\nabla \phi)^2 +c_{12} ]+
(s.t.)                \label{2.2}
\eeq
where $n$ is the parameter of dimensional regularization,
$C^2=C_{\mu\nu\alpha\beta }C^{\mu\nu\alpha\beta}$ is the square of Weyl
tensor and
$(\nabla\phi)^2=g^{\mu\nu}\;\nabla_\mu\phi\;\nabla_\nu\phi$. $"s.t."$
means "surface terms". The functions $c_{w,r,4,...,12}$ depend on
 $B(\phi), \lambda$ and on the derivatives of $B(\phi)$.
One can easily check  the surface
form of the other possible structures (see also \cite{eli,spec}).

We remark that the conformal-like invariance of the one-loop counterterms
requires an enormous cancellation of divergences. Let us consider, for
simplicity, the particular case of the conformal scalar theory $B(\phi) =
\frac{1}{12}\phi^2,\;A(\phi) = \frac{1}{2},\;B(\phi) = \lambda \phi^4$.
The conformal transformation has the form
$\phi\rightarrow\phi' = \phi e^{-\rho(x)},
\;\;g_{\mu\nu}\rightarrow g_{\mu\nu}' = g_{\mu\nu} e^{2\rho(x)}$.
It is fairly easy to see that even for the global transformation $\rho=const$
the expression (\ref{2.2})
is invariant only when $c_4=c_5=...=c_{11}=0$.
If one considers the local conformal
transformation, then it is necessary to have
$c_r=0$ as well. And so the one-loop counterterms preserve the symmetries of
the classical action if and only if they are given by the pure
Weyl term $\int d^4x \sqrt{-g} C_{\mu\nu\alpha\beta}^2$, and all other terms
are cancelled for any form of $B(\phi)$. As will be shown below it is not
the case.

Alternatively one can suppose that the transformation rule for the scalar
field is changed and in the counterterms it becomes inert to conformal
transformation. This can be achieved by the use of the special conformal
regularization \cite{etg,fv,bs} in a manner similar to the
last reference\footnote{It is not clear whether it is possible to do it
for arbitrary $B(\phi)$ }. In this case the invariance of (\ref{2.2})
under the global conformal transformation is fulfilled. The invariance under
local transformation requires $c_r=0$ and moreover other terms have to
compose the conformal invariant expressions established in \cite{a,eli}.
Note that the conformal regularization is not safe if we deal with the
quantum gravity theory. In fact it corresponds to some change
of variables in the path integral, that just leads to anomaly. Moreover,
this change of variables is nonlocal and therefore it can give contribution to
divergences \cite{a}. In any case we can trace the $R^2$ counterterm
which will certainly indicate to the violation of the conformal-like
symmetry in the one-loop
divergences, just as it happens in the Weyl conformal gravity.

Now we comment the relation between the different versions of our
theory with the conformal-like symmetry and General Relativity on
quantum level. In a general (nonsymmetric) metric-dilaton
theory (\ref{0.1}) the use of equations of motion
enables one to reduce the one-loop counterterms to the structures of
$c_w, c_r, c_7, c_{12}$ types only \cite{spec}. Then one can fine tune
three functions $A(\phi), B(\phi), C(\phi)$ and provide the on shell
renormalizability at one loop. It is not clear $ad \;hoc$ that it is
possible to make the same in a theory $S_{B(\phi),\lambda}$ with an
extra conformal-like symmetry. The symmetry results to that the equations
of motion are linearly dependent (\ref{1.13}). Therefore one can not
use those equations to cancel as much counterterms as in general
(nonsymmetric) theory (\ref{0.1}). Hence the equivalence of the theory
$S_{B(\phi),\lambda}$
with the General Relativity on quantum level
also requires the strong cancellation of the divergences. Moreover
the equivalence with the General Relativity contradicts to the
conformal-like invariance of the counterterms.
The one-loop calculation below is called to check whether any of those
cancellations really takes place or the conformal-like symmetry established
in the Theorem 2 is anomalous.

\vspace{5mm}

\noindent{\large \bf 4 One-loop calculation}
\vskip 1mm

In this section we shall present in some details the calculation of the
one-loop counterterms of the theory $S_{B(\phi),\lambda}$
with an arbitrary $B(\phi)$.
For our purposes we shall apply the background field method and the
Schwinger-De Witt technique \cite{DW} (see also \cite{book} for the
introduction).
The features of the metric-dilaton theory
leads to the necessity of some modifications of the calculational scheme,
basically developed in the similar two-dimensional theory \cite{odsh}
and recently applied to the general theory (\ref{0.1}).
The starting point of the calculations is the usual splitting of the fields
into background $g_{\mu\nu}, \phi$ and quantum $h_{\mu\nu}, \varphi$ ones
\beq
\phi \rightarrow \phi' = \varphi + \phi,              \;\;\;\;\;\;\;\;
\;   g_{\mu\nu} \rightarrow g'_{\mu\nu} + h_{\mu\nu}, \;\;\;\;\;\;\;\;
h_{\mu\nu} = {\bar h}_{\mu\nu}+\frac14 g_{\mu\nu}h,    \;\;\;\;\;\;\;\;
h=h_{\mu}^{\mu} \label{2.8}
\eeq
where we separated the trace and traceless
parts of the quantum metric for the sake of convenience. The one-loop
effective action is given by the standard general expression
\beq
\Gamma={i \over 2}\;\Tr\ln{\hat{H}}-i\;\Tr\ln {\hat{H}_{ghost}} \label{2.9}
\eeq
where $\hat{H}$ is the bilinear form of the action $S_{B(\phi),\lambda}$
 with added gauge fixing term and $\hat{H}_{ghost}$
is the bilinear form of the gauge ghosts action. Since the theory under
consideration is invariant under two --  diffeomorphism and
conformal-like symmetries, an additional gauge
condition is necessary to fix the last one. We shall follow Fradkin and
Tseytlin \cite{frts} who have derived the
counterterms in Weyl gravity and introduce this condition in the form
$h = 0$. Some note is in order. The condition $h = 0$ does not touch
the conformal-like invariance in the sector of background fields.
However this invariance is violated by the covariant gauge fixing term
\beq
S_{gf} = \int d^4 x \sqrt{-g}\;\chi_{\mu}\;\frac{\alpha}{2}\;\chi^{\mu},
\;\;\;\;\;\;\;\;\;\;\;
\chi_{\mu} = \nabla_{\alpha} \bar{h}_{\mu}^{\,\alpha}+
\beta \nabla_{\mu} \varphi                  \label{2.10}
\eeq
where $\alpha, \beta$ are some functions of the background dilaton,
which can be fine tuned to make the calculations more compact.
For instance, if one choose these functions as follows
\beq
\alpha=-B\;\;,\;\;\;\;\beta=-\frac{B_1}{B}
                                                        \label{2.11}
\eeq
then the bilinear part of the action $S+S_{gf}$ and the operator $\hat{H}$
has especially simple (minimal) structure
$$
\left(S + S_{gf}\right)^{(2)}
=\int d^4 x \sqrt{-g}\; {\omega} \hat{H} {\omega}^T
$$
\beq
\hat{H}=\hat{K}\Box+\hat{L}_{\rho}\nabla^{\rho}+\hat{M}    \label{2.12}
\eeq
Here $\omega=(\;\bar{h}_{\mu\nu},\; \varphi\; ),\;\;$ $T$ means
transposition,
$$
\hat{K}=\left(
\begin{array}{ccc}
\frac{B}{4}\left[\delta^{\mu\nu ,\alpha\beta}-
\frac{1}{4}\;g^{\mu\nu}\,g^{\alpha\beta}\right]
& 0\\
   0                                         & - \frac{B_1^2}{B}
\end{array}
\right)
$$
and the components of $\hat{L}_\rho$  and $\hat{M}$
can be easily extracted from the similar expression \cite{spec}
for the general theory (\ref{0.1}) with the help of condition $h=0$.

To separate the divergent part of $\Tr\ln\hat{H}$ we rewrite this
trace in a following way.
\beq
\Tr \ln\hat{H}  =\Tr\ln\hat{K}+
\Tr\ln\left(\hat{1}\Box + \hat{K}^{-1} \hat{L}^{\mu}\nabla_\mu
+\hat{K}^{-1}\hat{M} \right)                            \label{2.15}
\eeq
The first term does not give contribution to
the divergences whereas the second term has standard
minimal form and can be easily estimated with the use of
Schwinger-DeWitt method.
The bilinear form of the ghost action also has the minimal form
\beq
\hat{H}_{ghost}=
g^{\mu \alpha}\Box+\gamma(\nabla^{\alpha}\phi)\nabla^{\mu}
+ \gamma (\nabla^{\mu} \nabla^{\alpha} \phi) + R^{\mu \alpha} \label{2.16}
\eeq
and it's contribution to
the divergences can be easily derived with the use of
the standard technique.

Summing up the divergences of both terms of eq.(\ref{2.9})
we find that the one-loop divergences
have the form (\ref{2.2}) that is in a full accord with the power counting
consideration. The coefficient functions $c$ have the form
$$
c_w=
-{{17}\over {120}}\;\;\;\;\;\;\;\;\;\;\;
c_r=
{5\over {24}} - {{t_{1}}\over {6\,{t^2}}} +
{{{{  t_{1}}^2}}\over {8\,{  t^4}}}
$$$$
c_4=
{{-708\,{  t^6} + 368\,{  t^4}\, t_{1} -
247\,{  t^2}\,{{  t_{1}}^2} +
45\,{{  t_{1}}^3}}\over {96\,{  t^4}}} +
  {{\left( {  t^2} + 3\,  t_{1} \right) \,
t_{2}}\over {12\,{  t^3}}}
$$$$
c_5=
{{7\, t}\over 4} - {{11\,  t_{1}}\over {12\, t}} +
{{{{ t_{1}}^2}}\over {{ t^3}}},\;\;\;\;\;\;\;\;\;\;\;\;\;\;
c_6=
{{5\,{ t^2}}\over 2} - 3\, t_{1},\;\;\;\;\;\;\;\;\;\;\;\;\;\;
c_7=
{{9\,{\rm C}}\over 2B} +
\frac{{\rm C}_{2}}{B}\,\left( {1\over {6\,{  t^2}}} -
{{  t_{1}}\over {4\,{  t^4}}} \right)
$$$$
c_8=
{ { 1528{ t^8}
- 5584{  t^6} t_{1}
+ 3804{ t^4}{{  t_{1}}^2}
- 1804{  t^2}{{ t_{1}}^3}
+ 97{{t_{1}}^4} }
\over {256\,{ t^4}}}
+ { \left( 58 t^4
      + 16 t^2\,  t_1
      + 19{t_1}^2 + 2 t t_2
      \right) t_2
\over {32\,t^3}  }
$$$$
c_9=
{{-816\,{ t^6} + 554\,{ t^4}\,  t_{1} -
297\,{  t^2}\,{{  t_{1}}^2} +
64\,{{  t_{1}}^3}}\over {32\,{ t^3}}} +
  {{\left( 12\,{ t^2} +
7\, t_{1} \right) \, t_{2}}\over {8\,{ t^2}}},\;\;\;\;\;\;\;\;\;\;\;\;
c_{10}=
{{27\,{ t^2}}\over 8} +
{{31\,{{  t_{1}}^2}}\over {16\,{  t^2}}}
$$$$
c_{11}=
\frac{{9 {\rm C}}
({t_{1}}^2 -32t^4 + 4t_1\,t^2)}{16\,B\, t^2}
-{{9 t{\rm C}_{1}}\over 2B}
+ \frac{{\rm C}_{2}
(24t^4+120t^2t_1-111{t_1}^2+24 t_2)}{32\,B\,t^4}
+ \frac{{\rm C}_{3}(2t_1-t^2)}{2\,B\,t^3}
$$
\beq
c_{12}=
{9\,{\rm C}^2 \over {2B^2}} +
{{{\rm C}_2}^2 \over {8\,B^2\,t^4}}
                                   \label{2.17}
\eeq
where we have denoted $t = \frac{B_1}{B}$ for brevity. It is remarkable
that the dimensionless divergences depend only on $t$. Moreover, $t$
enters in the denominators and hence we can not put $B(\phi)$ equal to
constant and so be back to General Relativity. The source of this is
that the transformation (\ref{2.15})
that we have used is singular at $t = 0$.

The counterterms (\ref{2.17}) are not invariant under the conformal-like
transformation of Theorem 2. In particular, if we start with the
conformal  metric-scalar theory with $B(\phi) = \frac1{12}\phi^2$,
then (\ref{2.17}) differs from conformal invariant dilaton action
constructed in \cite{a,eli} and this  difference can not be removed
by the transformations from Theorems 1,2.
Let us now make some comments concerning the meaning of the $t$
dependence and the lack of conformal invariance.
On classical level the model under consideration is conformally
equivalent to General Relativity. On quantum level, within the background
field method that we are using here, the difference consists in:

i) Change of quantum variables related with conformal transformation and
with the consequent separation of fields into background and quantum parts.

ii) Conformal transformation and reparametrization of the background fields.
Since we have only two arbitrary functions related with the last point,
it is not possible to simplify (\ref{2.17}) and thus provide the
independence of $t$.

The crucial question is: whether the on shell one-loop
finiteness of General Relativity is lost in a new frame? The positive
answer indicate to the anomaly which break the conformal-like symmetry
on quantum level and gives the nontrivial contribution to the one-loop
counterterms. The detailed analysis of the on shell one-loop divergences
in a general nonsymmetric theory (\ref{0.1}) have been presented
in \cite{spec}.
Therefore we can discuss only the features of the conformal model.
In conformal theory the amount of the independent equations of motion
is less than in general one because of identity (\ref{1.13}). That is why
the equations of motion are insufficient to remove all the $\phi$
dependent counterterms in contrast to \cite{spec}. One can remove, for
instance, the $c_5, c_6, c_8, c_{10}$ type structures with the help of
classical equations of motion. However the $c_w, c_r, c_4, c_8, c_{11}$
ones remain and violate the on shell one-loop finiteness
 \footnote{We remark that the dimensional coupling function $C(\phi)$
and corresponding counterterms are not relevant here.}.

\vspace{5mm}

\noindent{\large \bf 5 Conclusion}
\vskip 1mm

We have considered the special class of the metric-dilaton theories
(\ref{0.1}) which satisfy the conditions (\ref{1.10}). In the Theorem 1 we
have proved
that all the actions  $S_{B(\phi),\lambda}$
of this class are converted into each other under
the
local conformal transformation of the metric field only. Two known actions --
namely the Einstein-Hilbert action of General Relativity and the action of
conformal scalar field belong to this class
and therefore all the  $S_{B(\phi),\lambda}$ theories
 can be regarded as different
frame for the description of Einstein gravity with (or without) cosmological
constant.
In the Theorem 2 we
have shown that all the $S_{B(\phi),\lambda}$ possess some
conformal-like symmetry including an arbitrary reparametrization of the
 scalar field supplemented by local conformal transformation of the metric.
On quantum level this symmetry is disturbed by anomaly.
The new anomalous degree of freedom
starts to propagate because of quantum effects. Since the theory under
consideration in not renormalizable by power counting, the anomalous
contributions enlarge the amount of the divergent structures and
finally the theories  $S_{B(\phi),0}$ with the nonconstant
$B$ are not finite on shell
whereas General Relativity is. Our result indicates to the
conformal noninvariance
of the $Diff$-invariant measure of the
path integral in four dimensional space-time
that was established earlier for the higher derivative Weyl gravity
\cite{frts,a}. Just as in the last case the conformal anomaly in the
theory  $S_{B(\phi),\lambda}$ gives contributions to the one-loop
divergences. Since one can conclude that
the effects of conformal anomaly in quantum gravity are essentially stronger
as compared with the conformal theories of matter fields in an external
gravitational field (one can see,
for example, the review of Duff \cite{20let} for the references on the
subject). In particular, the anomaly leads to that the parametrization
of gravity which we are using here does not satisfy the  general
theorem of Tyutin on the parametrization dependence of the
effective action in quantum field theory \cite{tyutin}.
The contribution of Jacobian of
the conformal transformation breaks the symmetries of the theory.
We remark that such things never happens with the gauge parameters
dependence which one can keep under control  at quantum level \cite{vlt}.

Not so far ago the conformal frame and conformal transformations
has been investigated in $2$ and $2+\varepsilon$  dimensional quantum gravity
\cite{DDK,AKKN,KST}. In that dimension the Einstein gravity and also the
metric-dilaton model (\ref{0.1} are renormalizable by power counting.
That is why the conformal anomaly does not lead to the nonrenormalizability
in this $d=2$. Thus one can regard our work as some investigation of
the difference between quantum gravity theories in two and four dimensions.

\vspace{5mm}

\noindent{\large \bf Acknowledgments}
\vskip 1mm

One of the authors (ILS) is grateful to M. Asorey, J.L.Buchbinder,
H. Kawai and I.V. Tyutin for the  discussions of the conformal anomaly
problem in quantum gravity.
ILS also thanks the Department of Physics at Hiroshima University
and Departamento de Fisica Teorica at
Universidad de Zaragoza  for warm hospitality. The  work of ILS has been
supported  in part by the RFFR (Russia), project
no. 94-02-03234, and by ISF (Soros Foundation), grant RI1000.

\vspace{5mm}

\noindent{\large \bf Appendix}
\vskip 1mm

In this Appendix we consider an interesting symmetry property
which takes place in the theory with the action
$$
S\left[g_{\mu\nu}; B(\phi), {\cal B}(\phi); \lambda, \tau \right]
=S_{B(\phi),\lambda}
+ S_{{\cal B}(\phi),\tau}    \eqno{(A1)}
$$
where we use our notation
$$
S_{B(\phi),\lambda} = \int d^4x \sqrt{-g}
\; \left\{ A(\phi)g^{\mu\nu}
\partial_{\mu}\phi
\partial_{\nu}\phi + B(\phi)R + \lambda { B}^2  \right\}  \eqno{(A2)}
$$
with ${A} = A[B] = \frac{3 {B_1}^2}{B} = 0$ and the same for
$S_{{\cal B}(\phi),\tau}$. Indeed the theory (A1) belongs to general
nonsymmetric class of the models (\ref{0.1})
rather than to the models with conformal-like symmetry. However this
theory manifests a very interesting property under the transformation of
Theorem 1.

At first we shall see how the general
action (A1) behaves under the conformal transformation of the metric
$$
g_{\mu\nu} = {\bar g}_{\mu\nu}\;e^{2\sigma(\phi)}        \eqno{(A3)}
$$
In four dimensions the geometrical quantities transform as
$$
\sqrt{-g} = \sqrt{-{\bar g}}e^{4\sigma(\phi)},
\;\;\;\;\;\;\;\;\;g^{\mu\nu}= {\bar g}^{\mu\nu}e^{- 2\sigma(\phi)}
$$$$
R = e^{- 2\sigma(\phi)}
\left[{\bar R} - 6{\bar \Box}\sigma -6
({\bar \nabla}_\mu\sigma)( {\bar \nabla}^\mu\sigma) \right] \eqno{(A4)}
$$
Substituting (A4) into (A2) and then to (A1) we find the following
transformation rule for the last one.
$$
S\left[g_{\mu\nu}; B(\phi), {\cal B}(\phi); \lambda, \tau \right]
=S\left[{\bar g}_{\mu\nu}; B(\phi)e^{- 2\sigma(\phi)},
{\cal B}(\phi)e^{- 2\sigma(\phi)}; \lambda, \tau \right]      \eqno{(A5)}
$$

Now we can explore an interesting particular case of the theory (A1)
and transformation (A3). Let both things are chose in such a
manner that $B(\phi) = \gamma = const $ and
also ${\cal B}(\phi)e^{2\sigma(\phi)} = {\bar \gamma}
= const $. The last condition immediately gives $e^{2\sigma(\phi)} =
\frac{{\bar \gamma}}{{\cal B}(\phi)}$.  Then (A5) becomes
$$
S\left[g_{\mu\nu}; \gamma, {\cal B}(\phi); \lambda, \tau \right] =
S\left[{\bar g}_{\mu\nu}\;;
\frac{{\bar \gamma} \gamma}{{\cal B}(\phi)}, {\bar \gamma}
\;; \lambda, \tau \right]                            \eqno{(A6)}
$$
If one put ${\bar \gamma} = \gamma^{- 1}$ then the last equation takes
especially simple form
$$
S\left[g_{\mu\nu}; \gamma, {\cal B}(\phi); \lambda, \tau \right] =
S\left[ {\bar g}_{\mu\nu}\;;
\frac{1}{\gamma}, \frac{1}{{\cal B}(\phi)}\;;
\lambda, \tau \right]                              \eqno{(A7)}
$$
that deserve to be written in an explicit form
$$
\int d^4x \sqrt{-g}
\; \left\{ \gamma R + \lambda \gamma^2
+ A[B]g^{\mu\nu}\partial_{\mu}\phi
\partial_{\nu}\phi + B(\phi)R + \tau { B}^2  \right\} =
$$$$
= \int d^4x \sqrt{-{\bar g}}
\; \left\{ \frac{1}{\gamma}{\bar R} + \frac{\tau}{\gamma^2}
+ A\left[\frac{1}{B}\right] {\bar g}^{\mu\nu}\partial_{\mu}\phi
\partial_{\nu}\phi + \frac{1}{B(\phi)}{\bar R} +
\frac{\lambda}{B^2(\phi)}  \right\}
  \eqno{(A8)}
$$
The above transformation has a dual form. Since it describes the
invertion of  the coupling constant $\gamma$ and
function $B(\phi)$ this transformation shows that there is some link
between strong and week coupling regimes in some of the models (\ref{0.1}).

The particular case of (A8) with $B(\phi)=\frac12\phi^2$ and
 $\lambda = \tau= 0$
corresponds to the second theorem of Bekenstein
\cite{5} whereas the first theorem of \cite{5} results from the conformal
equivalence of the different versions of (\ref{0.1}) which do not satisfy
(\ref{1.10}) (see also \cite{6} for the case  of the nonzero $\lambda,\tau$).

\end{document}